%% file: ml.tex
\newif\ifmulticol	\multicoltrue
\newif\ifshowgit	\showgittrue		% switches footer on/off
\newif\ifgitlocal	\gitlocalfalse		% use local file gitHeadLocal.gin
\newif\ifbiblatex	\biblatexfalse		% defaults to bibtex if false
\newif\ifbibnum		\bibnumtrue 		% num => superscripts, otherwise auth date
\newif\ifbibsort	\bibsortfalse		% biblatex num sort in order of occurrence
\newif\iflineno		\linenofalse
\newif\iftoc		\tocfalse

\newif\iflucida		\lucidatrue
\newif\ifcm			\cmfalse
\newif\ifcharter	\charterfalse		% use for arXiv, use xelatex
\newif\ifcharterotf	\charterotffalse	% requires xelatex

\newcommand*{\secnumdepth}{0}			% sets secnumdepth in \mymaketitle
\newcommand*{\tocdepth}{2}				% sets tocdepth in \mymaketitle
\newcommand*{\reftitle}{References}		% title for references section
\newcommand*{\mytitleskip}{-0.2in}		% change skip after title

\multicoltrue\showgittrue\gitlocaltrue\biblatexfalse\bibnumtrue\lucidafalse\chartertrue

\newcommand*{\mydocfontsize}{\ifcharter11pt\else10pt\fi}
\newcommand*{\setcol}{\ifmulticol twocolumn\else onecolumn\fi}

\documentclass[\mydocfontsize,\setcol]{article}

\newcommand*{\pdfauthor}{Steven A.~Frank}
\newcommand*{\pdftitle}{Circuit design in biology and machine learning. II. Anomaly detection}

\usepackage{afterpage}
\usepackage{enumitem}

\input ml.sty
\input ml-def

\hypersetup{linkcolor={Maroon4}, urlcolor={Maroon4}}

\newcommand*{\mytitle}{\pdftitle}

\newcommand*{\myauthor}{Steven A.\ Frank}
\newcommand*{\myaddress}{Department of Ecology and Evolutionary Biology, University of California, Irvine, CA 92697--2525, USA}
\newcommand*{\mycontact}{\href{https://stevefrank.org}{https://stevefrank.org}}

\newcommand*{\mykeywords}{Biological design, evolution, artificial intelligence, boosted decision trees, dimensional reduction, internal model principle}
\setabstract{-0.07}{\iftoc-.0\else0.0\fi}{%
Anomaly detection is a well-established field in machine learning, identifying observations that deviate from typical patterns. The principles of anomaly detection could enhance our understanding of how biological systems recognize and respond to atypical environmental inputs. However, this approach has received limited attention in analyses of cellular and physiological circuits. This study builds on machine learning techniques---such as dimensionality reduction, boosted decision trees, and anomaly classification---to develop a conceptual framework for biological circuits. One problem is that machine learning circuits tend to be unrealistically large for use by cellular and physiological systems. I therefore focus on minimal circuits inspired by machine learning concepts, reduced to cellular scale. Through illustrative models, I demonstrate that small circuits can provide useful classification of anomalies. The analysis also shows how principles from machine learning---such as temporal and atemporal anomaly detection, multivariate signal integration, and hierarchical decision-making cascades---can inform hypotheses about the design and evolution of cellular circuits. This interdisciplinary approach enhances our understanding of cellular circuits and highlights the universal nature of computational strategies across biological and artificial systems.

\vskip14pt

\textbf{Teaser text:} How do cells detect unusual changes in their environment? This study explores machine learning methods that identify anomalous patterns, illuminating the design and evolution of cellular circuits. Simple models inspired by machine learning demonstrate that even minimal biological systems can effectively recognize and classify uncommon signals. This link between how machines learn and how cells evolve introduces a new theoretical approach to the evolutionary design of cellular circuits, highlighting universal strategies shared between living organisms and computational systems.
}

\begin{document}

\mymaketitle

%% 1st parm is skip on left column at start of TOC, 2nd param is skip after TOC
\iftoc\mytoc{-24pt}{\newpage}\fi

\section{Introduction}

Many biological circuits sense danger. Some respond to common molecular patterns associated with attack. Others perceive environmental threats for which fear or fighting may be helpful \autocite{heil14danger,leroux15bacterial,nesse19the-smoke,klein21fear,moscarello22the-central}.

An unusual or surprising environment provides another clue of danger. For example, the absence of an expected event could signal an anomaly. The famous comment by Sherlock Holmes about the dog that did not bark illustrates an anomalous absence of an expected event \autocite{doyle93theadventure}.

A Scotland Yard detective asked Holmes: “Is there any other point to which you would wish to draw my attention?” Holmes answered: “To the curious incident of the dog in the night-time.” The detective replied: “The dog did nothing in the night-time.” Holmes countered: “That was the curious incident.”

Intuitively, humans have a sense of anomaly, when unexpected events trigger heightened alertness. The word ``eerie'' captures the notion of discomfort when ``things don't add up'' in an unfamiliar situation.

For these reasons, anomaly detection focuses on the deviation from what is typical. An anomaly detection circuit must learn an internal model of the typical pattern. Any departure from that model triggers a warning. This approach contrasts with detecting specific danger signals that directly indicate peril, emphasizing instead deviations from common observations.

In mammalian brains, hippocampal circuits detect anomalies \autocite{kumaran06anunexpected,kumaran07match--mismatch,bhasin22dynamic}. Immune systems may have such circuits \autocite{dasgupta99an-anomaly,ramadan17editorial:}. Self versus nonself recognition is not fully understood \autocite{koncz24a-journey} and might, in some cases, depend on detecting anomalous patterns as nonself. Plants might use anomalous volatile organic compounds of neighbors as nonspecific danger signals \autocite{meents20plant--plant,duc22volatile}. However, few biological studies emphasize nonspecific anomaly detection.

This article introduces anomaly detection in machine learning \autocite{omar13machine,nassif21machine,pang21deep,ruff21aunifying}. Computational models use a wide variety of circuit types to detect anomalies. Those different types of computational circuits suggest the kinds of biological circuits that might detect anomalies. Because anomaly detection is a type of classification problem, aspects of this topic also provide insight into other biological challenges of classification.

\section{Contributions of this work}

\subsection{Overview of the series}

This article continues the series on circuit design in biology and machine learning \autocite{frank24circuit}. The series uses insights from machine learning to understand how evolutionary processes build biological circuits. The first article in the series introduced the motivation and challenges for linking biological and machine learning circuits, with examples \autocite{frank24circuit}. This subsection adds further background.

Three facts suggest that machine learning may provide insight into the evolutionary aspects of biological design. First, machine learning and biological organisms often face similar challenges. How can environmental inputs be classified into categories? How can a system predict future inputs? What is the best response to a type of environment?

Second, natural selection is one type of learning algorithm. Machine learning deploys a broader range of algorithms. But those different algorithms tend to modify systems in broadly similar ways \autocite{de-jong06evolutionary,holland75adaptation,holland95hidden}.

Third, machine learning and biology often solve challenges by using a computational network to build an input-output response circuit. Here, we think of a biochemical network as a kind of circuit that takes inputs and computes outputs. When machine learning computes solutions without an explicit network, usually the computation can be embedded within a network to achieve the same result.

The fact that machine learning and biology typically build responses by creating computational circuits means that we can study how machine learning solves particular kinds of problems and use those solutions to make predictions about how evolutionary processes design biological circuits to solve the same sorts of challenges.

This series emphasizes simple biochemical circuits, primarily in cells. The analogy between neurobiological and machine learning circuits is well known, although directly linking the architecture and function of biological and computational circuits remains an ongoing challenge \autocite{mcculloch43a-logical,hebb49theorganization,rumelhart86parallel,jiahui23modeling,kanwisher23using}.

By contrast, relatively little work has been done to match cellular or physiological circuits to common machine learning architectures. Two challenges arise.

First, although many biochemical circuits in cells have been identified and partially understood, it is not easy to describe complete circuits, understand their computational architecture, and evaluate the sorts of computations that are used to achieve their function.

Second, computational networks in machine learning tend to be much larger than could reasonably fit within a cell. Thus, we must develop new machine learning models that emphasize greatly reduced size.

Given those constraints, this series primarily aims to sketch the outlines for a new theory that links these two subjects. Some general predictions arise about the architecture of biological circuits. Overall, these articles show the broad conceptual links between particular external challenges and the types of biological circuits that may be favored by evolutionary processes.

\subsection{Insights from anomaly detection}

This article develops the following points, often with simple illustrative models and example quantitative analyses.
\begin{itemize}[left=0cm, labelsep=0.3cm]

\item Machine learning provides new ideas for how cellular and physiological circuits may solve anomaly detection.

\item Some challenges require evaluating a single atemporal multivariate input for anomaly. Others require estimating deviations from recent temporal trends. Simple models illustrate different circuit designs for atemporal and temporal cases.

\item Detecting anomalies often requires evaluating multivariate patterns in inputs by integrating signals from ensembles of sensors or receptors. This article reviews basic measures of signal information.

\item Digital sensors reduce continuous analog inputs to discrete binary outputs, losing information but also reducing sensitivity to noise and measurement error. Digital sensors are easier to implement and easier to combine into broader circuits.

\item Machine learning uses big circuits. Cells require small circuits. This article shows that small circuits can achieve significant resolving power.

\item Some anomalies differ in mean input values. Summing the inferences by individual sensor outputs provides a good response.

\item Other anomalies differ in correlations between inputs. Decision trees work well, each sensor responding within a sequence based on the output of prior sensors.

\item Machine learning often deploys cascades of circuits, such as a cascade of separate decision trees.

\item Each small circuit passes its response to the next circuit, which corrects errors and boosts response quality.

\item Learning a sequence of boosted circuits matches the likely way that evolution works, sequentially improving an existing cascade of small modular subsolutions.

\item Dimensional reduction provides a potential alternative for anomaly detection. Typical multivariate inputs can be reduced to a lower dimension, similar to principal components analysis. An anomalous input tends to be relatively distant from typical inputs in the reduced space.

\item Small encoder circuits can reduce dimensionality, classifying differences in the correlational structure of typical and anomalous inputs. In general, dimensional reduction is likely to be a major feature of biological circuits.

\item As in all problems of biological design, evolutionary tuning with respect to tradeoffs inevitably plays a central role in shaping biological circuits.

\end{itemize}

\section{Timescale}

\subsection{Instantaneous versus time-dependent inputs}

Timescale broadly influences the kinds of circuits that can succeed in anomaly detection. Most anomaly detection methods consider multiple inputs at one point in time. If a single multivariate input is unusual compared with the set of typical multivariate points, then that unusual input is classified as an anomaly.

In some cases, an anomaly must be considered with respect to recent temporal trends \autocite{blazquez-garcia21areview,choi21deep}. For example, reactive oxygen species are often used as weapons in microbial warfare. A rapid increase in concentration of these dangerously reactive molecules may signal attack.

For multivariate problems that use a single atemporal input, a machine learning method typically classifies by some sort of clustering, partitioning, or dimensional reduction \autocite{omar13machine,nassif21machine,pang21deep,ruff21aunifying}. The common inputs fall toward one cluster, or in a particular direction away from a partition, or in a particular location in a reduced space of constructed dimensions. The anomalous inputs are those that are not near the common set.

Temporal problems also require classification \autocite{blazquez-garcia21areview,choi21deep}. However, before classification, one must adjust for the temporal dependence of the input stream. For example, typical inputs may follow a rising trend. An anomaly must be measured against the expected input from the current trend, which requires a circuit to maintain an updated trend estimate.

\subsection{Biological response times}

Atemporal classification of anomalies demands a sufficiently fast circuit. The multivariate perception of input must be accomplished before the environment changes significantly. The calculations to classify must follow with sufficiently short lag to allow an appropriate response. 

A neurobiological circuit would likely be quick enough to do atemporal classification. For cellular or physiological circuits, response speeds vary widely for different components, from slow biochemical reactions to fast receptors. If the environment changes significantly faster than a circuit's classification inference, then such a circuit may not be able to classify the current environment as if it were an instantaneous isolated event.

Temporal classification over input sequences alters the timescale constraints. The circuit's estimate of trends in inputs may update continuously, although with a time lag. The circuit can work well if its update lag is shorter than the timescale over which environmental trends change.

For temporal classification problems, neurobiological circuits would likely be quick enough for most challenges. Cellular and physiological circuits may sometimes be quick enough if intrinsic temporal smoothing of trend estimation provides sufficient information.

At present, we know little about the cellular and physiological response times of anomaly detection circuits. I limit discussion to three brief comments. 

First, cellular receptors can potentially respond on the timescale of their ligand on-off rates, which are often very fast. So, at the receptor level, sensory information may be able to keep up with environmental change.

Second, some cellular states depend on electric gradients, which change rapidly and can be transmitted at relatively high speed \autocite{levin14molecular}. These fast components of cellular response might provide a sufficient basis for speedy circuits.

Third, slower downstream biochemical reactions might constrain circuit design. Different biochemical processes vary in their response times \autocite{fell97understanding}. Altering the concentrations of reactants often triggers the fastest response. Covalent modifications of enzymes are slower than changes in reactant concentrations.  Altering enzyme production or degradation rates is typically the slowest modification of biochemical circuits.

Many other factors could change biochemical response times. However, those factors are likely to be slow relative to the responses of receptors or electric gradients.

\section{Simple mechanisms}

\subsection{Atemporal biochemical mechanism}

This subsection briefly illustrates how we may think about mechanistic components in biological circuits. The example describes a simple circuit for atemporal challenges. The following subsection considers temporal challenges. 

As a first step, a circuit may determine how each input dimension deviates from its typical value. We begin with a widely observed empirical relation in biochemistry, the Hill function \autocite{frank13input-output,zhang13ultrasensitive,martinez-corral24the-hill}. This function describes a common pattern by which an input level is transformed into an output response as
\begin{equation}\label{eq:hill}
  h(u|m,k)=\frac{u^k}{m^k+u^k},
\end{equation}
in which $u$ is the input, $k$ is a coefficient that determines shape, and $m$ is the input at which the response achieves one-half of its maximum level. The notation $u|m,k$ identifies $u$ as the input variable given the parameters $m$ and $k$ that determine response. In the following, I assume that all functions share the same $k$ value, which is dropped from the notation to simplify the expressions.

\begin{figure*}[t]
\centering
\vskip-1.5cm
\includegraphics[width=0.5\hsize]{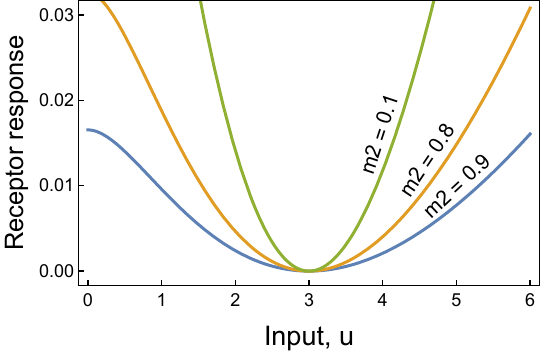}
\vskip0pt
\caption{Receptor response for atemporal anomaly detection. When the input is at the typical value of $u^*=3$, the receptor responds with a minimal value. As the input increasingly deviates from its typical value, the receptor returns an increasing response. The likelihood of an anomalous condition rises with the receptor response value. Thus, this receptor provides a simple atemporal way to classify inputs as normal or anomalous. This figure derives from \Eq{anomRecept}, with $m_1=1$, $k=2$, and $a$ set so that $u^*=3$ is a minimum. To get a minimum at $u^*$, we search for $a$ such that $\dd \hat{r}/\dd u=0$ and $\dd^2\hat{r}/\dd u^2>0$ when evaluated at $u^*$. For $k=2$ and $m_1>m_2$, we obtain $a = \left(2m_1^2+u^*+4\right) \left(m_2^2+u^*+2\right)^2{\Big/}\left(2 m_2^2+u^*+4\right)\left(m_1^2+u^*+2\right)^2.$ As $m_1$ and $m_2$ become large, $a\rightarrow (m_2/m_1)^2$.
}
\label{fig:anomaly}
\end{figure*}

We seek a circuit that identifies an anomalous input by its deviation from a standard input level, $u^*$. Suppose a receptor balances stimulative and repressive forces in relation to input level, given by the difference between two Hill functions
\begin{equation}\label{eq:hillbalance}
  \hat{r}(u|m_1,m_2) = h(u|m_1) - a\,h(u|m_2),
\end{equation}
in which $a$ is a weighting on the repressive effect to achieve a relative balance between the two forces. We can choose $a$ so that $\hat{r}$ evaluated at $u^*$ is a minimum, which gives a circuit that increases in output as the input, $u$, deviates from the minimum, $u^*$ (see caption for \Fig{anomaly}).

For numerical convenience, we can subtract the value of $\hat{r}$ at its minimum to obtain a receptor that returns zero when the input is at its standard level, $u^*$, and returns increasing values as $u$ increasingly deviates from $u^*$, as
\begin{equation}\label{eq:anomRecept}
  r(u|m_1,m_2) = \hat{r}(u|m_1,m_2) - \hat{r}(u^*|m_1,m_2).
\end{equation}
\Figure{anomaly} illustrates how this receptor identifies anomalous deviations from typical input values. The cutoff for classifying an anomaly may evolve by natural selection in a biological context or be estimated from data in a machine learning context. Multivariate input requires combining multiple receptor responses to identify anomalies.

\subsection{Anomalous deviation from temporal trend}

Suppose the value of typical inputs, $u^*$, changes over time. A circuit must estimate the current typical input. That estimate of $u^*$ and the current input value, $u$, can be used in \Eq{anomRecept} to get the receptor signal.

If we choose sufficiently large values of the $m$ parameters, then $a\rightarrow (m_2/m_1)^2$ (see caption of \Fig{anomaly}). With approximately constant $a$, to calculate the receptor response in \Eq{anomRecept}, we only need to track the dynamics of $u^*$ given the input, $u$. We can track the dynamics of $u^*$ by including in the circuit
\begin{equation}\label{eq:anomtrend}
  \dot{u}^* = \Gl\lr{u-u^*},
\end{equation}
in which $u$ is a stochastically changing input, and $u^*$ is an exponential moving average of $u$, with the overdot denoting the derivative with respect to time. The parameter $\Gl$ controls the speed at which the internal moving average estimate responds to changing inputs. This process simply describes the production or degradation rate of a molecule in response to the level of a stimulating input.

\begin{figure*}[t]
\centering
%\vskip-1.5cm
\includegraphics[width=0.95\hsize]{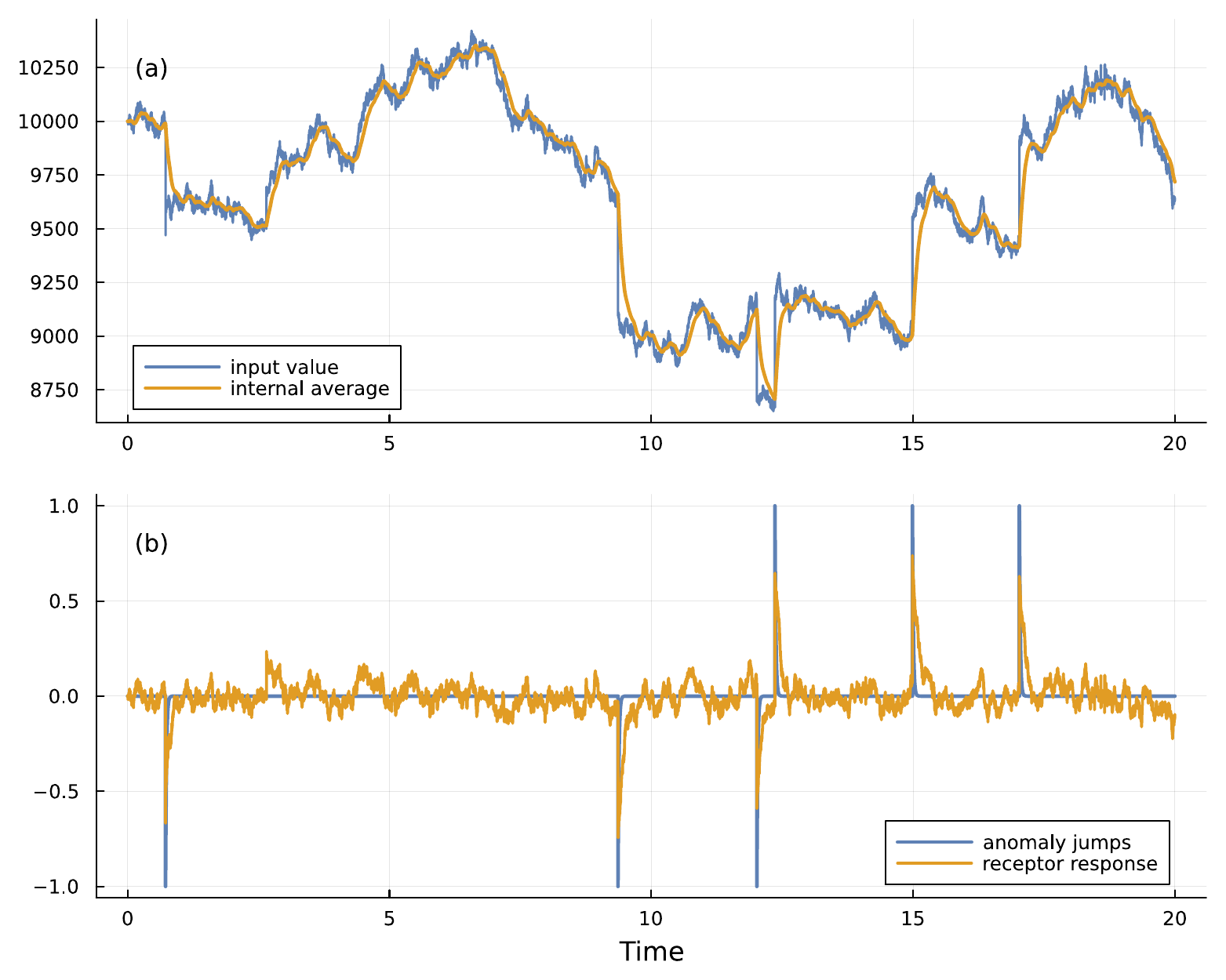}
\vskip0pt
\caption{Receptor response for temporal anomaly detection. (a) The blue input signal, $u$, was generated by a stochastic process $\dd u = 0.0002(10000-u)\,\dd t + 0.02u\,\dd W + zu\,\dd N$, in which $\dd W$ is a Wiener process that generates continuous Gaussian noise with a mean of $0$ and a standard deviation of $1$, and $\dd N$ is a Poisson jump process that generates random discrete jumps at rate $0.2$. Each jump multiplies the current input, $u$, by $z$, which for each jump takes on a value $0.95$ or $1.05$ with equal probability. The gold moving average, $u^*$, is given by \Eq{anomtrend} with $\Gl=10$. (b) The blue spikes show the timing and direction of the random anomalous jumps for this sample run. The levels of $\pm1$ for the spikes are arbitrary values. The gold curve shows the receptor output from \Eq{anomRecept} multiplied by $25$, with $k=2$, $m_1=10,000$, $m_2=1000$, and $a$ given by the solution in the caption for \Fig{anomaly} with $u^*=m_1$. The gold receptor spikes match the blue anomalous input jumps, signaling anomalies.  The freely available Julia computer code provides full details about assumptions and methods for all figures in this article \autocite{frank24circuit-codeb}.
}
\label{fig:anomaly2}
\end{figure*}

Using the moving average estimate for typical inputs, $u^*$, in \Eq{anomRecept} allows the receptor to adjust to changing environmental conditions. \Figure{anomaly2} shows how this adaptable receptor detects anomalous deviations in the input signal.

In the only similar model that I found, a circuit estimates the ratio of a current input relative to the recent value of typical inputs \autocite{goentoro09the-incoherent}. The goal was ratio estimation, which the authors called \textit{fold change}. Significant deviations of the ratio from one could also be used for anomaly detection.

To obtain a fold-change circuit, instead of using the receptor in \Eq{anomRecept}, we combine the moving average estimate from \Eq{anomtrend} with
\begin{equation}\label{eq:foldchange}
  \dot{y} = \Gg\lr{\frac{u}{u^*}-y},
\end{equation}
which approximately matches the circuit given published fold-change circuit \autocite{goentoro09the-incoherent}. If the dynamics respond sufficiently quickly to changes, with large enough $\Gl$ and $\Gg$, then $y$ measures the ratio of the current input $u$ to the recent typical input, $u^*$.

In this ratio-estimating biochemical circuit, the response time may differ from the response time of a receptor-based circuit approximated by \Eq{anomRecept}. Typically, receptors respond more quickly than biochemical reactions. However, it is not clear if natural processes more easily build and tune circuits based on receptors or biochemical reactions in solution.

\section{Multivariate signals}

The prior subsections analyzed deviations in a single dimension. Detecting anomalies often requires combining information from multiple dimensions. For example, identifying attacks on a computer network depends on the number of data bytes sent to the target computer that may be under attack, the number of data bytes returned to the potential attacker by the target computer, and the type of connection such as email or web page.

Two widely used test datasets for computer network attack include those network measures along with several other dimensions of data \autocite{ring22cidds-001,tavallaee22nsl-kdd}. The challenge is to classify whether a network connection to a target computer is a normal use or an attack. Is the connection pattern described by the multivariate measures of the connection typical or anomalous?

Many different machine learning methods have been applied to these benchmark datasets \autocite{thapa20comparison,salih21evaluation,torabi23practical}. The next subsection begins by evaluating each data dimension independently to infer anomalies and then combining the information in the independent dimensions.

The following subsections include the information in the correlation between dimensions by combining the dimensions into a decision tree or by using hierarchical dimension reduction by encoders, two widely used machine learning methods that may map relatively easily to biological circuits.

I start with artificial data to illustrate the methods. I then turn to real data that contrasts typical computer network connections with anomalous connections from attack.

I use the computer data because we do not have large datasets with multivariate measurements of typical and anomalous biological inputs. The goal here is to illustrate the key principles of circuit design that may be important for understanding how natural processes shape biological responses. Anomaly detection has hardly been studied in cellular biology but seems likely to be important in some circumstances.

\subsection{Independent data dimensions and ensembles}

Suppose an input generates $n$ independent data dimensions. For a typical input, the value in each dimension is a random sample from a normal distribution with mean $m_t$ and standard deviation $\Gs$. Similarly, an anomalous input generates $n$ independent values, each sampled from a normal distribution with mean $m_a$ and standard deviation $\Gs$. Assume typical inputs are usually smaller than anomalous inputs, $m_t<m_a$.

Suppose a biological circuit can average the $n$ independent values associated with each input. Then the standard deviation of the average value is $\Gs/\sqrt{n}$. The circuit classifies the average value as typical if it is less than a threshold value, $\Gt$, and anomalous if greater than the threshold.

\begin{figure*}[t]
\centering
%\vskip-1.5cm
\includegraphics[width=0.9\hsize]{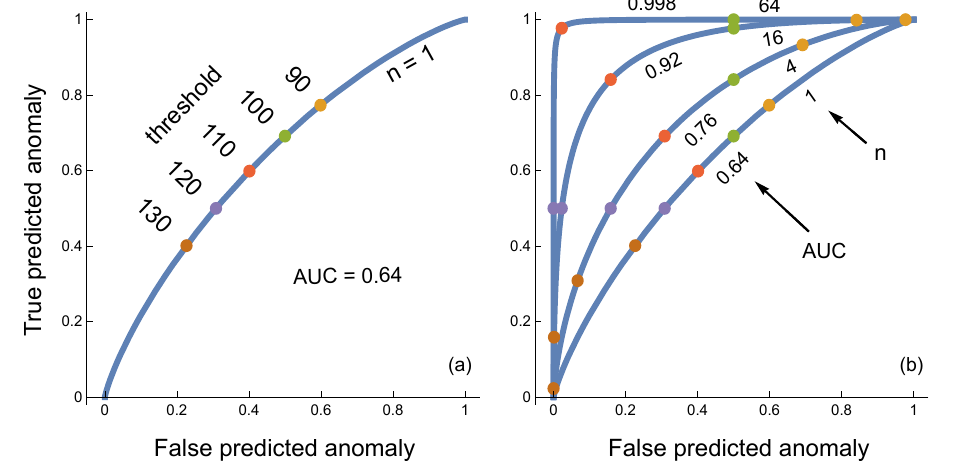}
\vskip0pt
\caption{Classification of input as typical or anomalous by a circuit that averages $n$ independent input values and makes a decision based on the average value. Inputs are continuous numerical values. In this example, I generated inputs by randomly sampling a normal distribution with a standard deviation of $\Gs=40$. For typical and anomalous inputs, the distribution means are $100$ and $120$, respectively. (a) The circuit takes $n=1$ dimensions of input. The circuit uses a threshold, $\Gt$, such that the circuit classifies inputs below the threshold as typical and above the threshold as anomalous. The curve plots the frequency of truly predicted anomalies as a function of $\Gt$ versus the frequency of falsely predicted anomalies as a function of $\Gt$, generating a receiver operating characteristic (ROC) curve. The area under the curve (AUC) measures the resolving power of the circuit that describes the tradeoff between true positive and false positive classifications over all of the thresholds. (b) For each case in which the true generating process is either typical or anomalous, I generated $n$ independent samples for the associated probability distribution. The circuit measures the average of the inputs, which, when compared to the $n=1$ case in the left panel, has the same mean and a reduced standard deviation, $\Gs/\sqrt{n}=40/\sqrt{n}$. The reduced variation provides the circuit with greater resolving power, described by the increasing AUC with increasing $n$.
}
\label{fig:oneR}
\end{figure*}

\Figure{oneR}a illustrates how a change in threshold value alters the circuit's success at classifying inputs. A smaller threshold causes a higher rate of classification as anomalous, which increases both the true rate of predicting anomalies and the false rate of predicting anomalies. As the threshold changes, the curve traces the tradeoffs between those different aspects of successful classification. The area under the curve (AUC) provides one way to measure the overall quality of the circuit's ability to classify inputs.

\Figure{oneR}b shows how the circuit's response characteristics improve for increasing levels of $n$, the number of data dimensions sampled by the circuit. More data dimensions provide more precise information about whether the input is typical or anomalous.

\subsection{Digital circuits}

Precise estimates for each of the $n$ data values may be difficult for biological sensors, making the circuit sensitive to perturbations in measurement. Suppose instead that each sensor encoded its response in a binary way, which we can label as $0$ or $1$. In other words, each sensor converts its analog input to a digital output: a $0$ response when the value is below some threshold and a $1$ response above the threshold. Such analog to digital conversion can be approximated by the Hill function response described by \Eq{hill}, which is widely observed in biology \autocite{frank13input-output,zhang13ultrasensitive,martinez-corral24the-hill}.

With digital sensors, a circuit only has to combine the information into an overall frequency of $1$ values, which are the anomaly signals. For example, if each sensor can trigger the activation of a transcription factor, then those transcription factors can bind to a gene promoter. By this process, the promoter can produce a response that grades with the overall frequency of anomaly signals from the sensors.

This digital circuit requires two threshold values. First, $\Gt$ sets the point below which an individual sensor returns $0$ for a typical input and above which the sensor returns $1$ for an anomalous input. Second, a threshold $\Gf$ sets the frequency of $1$ responses among the individual sensors required for the circuit to return an overall classification of anomalous for a multivariate input.

\begin{figure*}[t]
\centering
%\vskip-1.5cm
\includegraphics[width=0.9\hsize]{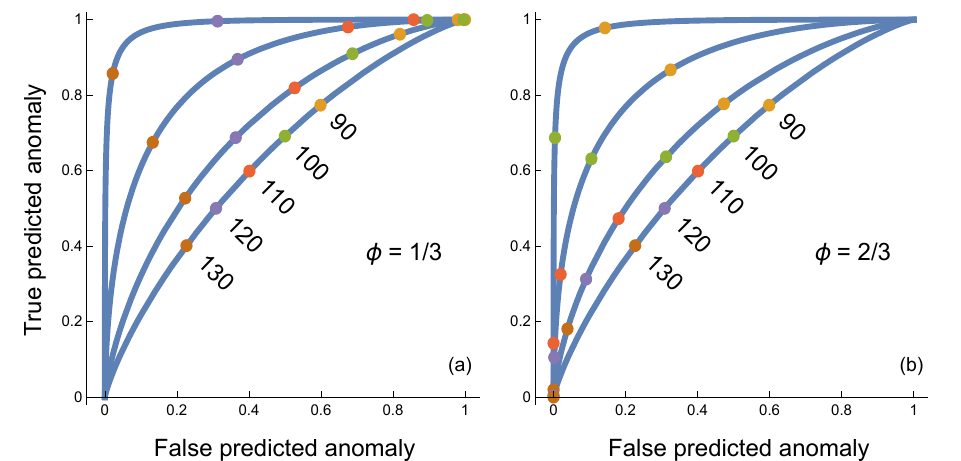}
\vskip0pt
\caption{Anomaly classifier in which each sensor does an analog to digital conversion, with $n=1,4,16, 64$ sensors for curves from bottom to top. The generation of continuous input into each sensor is described in the caption of \Fig{oneR}. In this case, each sensor receives an independent input value and independently scores its input as $0$ for typical or $1$ for anomalous based on the threshold shared by all sensors. Colored circles on each curve denote particular threshold values for the individual sensors. The overall classification by the circuit depends on the frequency of $1$ values returned by the individual sensors. The circuit returns an anomaly if the frequency of $1$ values by individual sensors is greater than $\mathrm{ceiling}(\Gf n)/n$, in which the ceiling function returns the smallest integer greater than or equal to its argument. (a) Curves for $\Gf=1/3$. (b) Curves for $\Gf=2/3$. Increasing the frequency threshold, $\Gf$, lowers both the true and false positive classification rates, which can be seen by comparing the same sensor threshold values between the two panels. When $\Gf=1/2$, the threshold locations, $\Gt$, are intermediate between the two panels. The AUCs are $0.64,0.72,0.87,0.99$ for curves from bottom to top in both panels. The AUC values for $\Gf=1/2$ are slightly higher in the third significant digit for larger $n$. Overall, the AUC circuit performance is very flat as a function of varying frequency cutoff, $\Gf$, over $(1/3,2/3)$, suggesting that $\Gf$ may be a nearly neutral trait over a wide range in the AUC sense of measuring performance over a range of individual sensor thresholds, $\Gt$.
}
\label{fig:oneR2}
\end{figure*}

\Figure{oneR2} shows how the two thresholds interact. Higher curves correspond to increasing numbers of sensors, $n$. In (a), with $\Gf=1/3$, low thresholds for the individual sensors, $\Gt$, cause increasing $n$ to provide relatively high false predicted anomalies (false positives). This pattern can be seen by starting with the lower curve for $n=1$ and the smallest labeled threshold of 90 marked by the gold circle.

As $n$ increases and the curves rise, the gold circle for 90 moves to the right because the rate of false predicted anomalies along the $x$-axis increases. The reason is that with both a low individual sensor threshold and a low overall threshold, the expected outcome for a typical input is a false positive prediction of anomaly. As $n$ increases, the variance declines and the expected outcome increasingly dominates.

In \Fig{oneR2}b, with $\Gf=2/3$, high thresholds for the individual sensors tend to cause an increase in predictions of typical for the inputs. More predictions of typical raise the false positive rate of typical predictions, which corresponds to a reduced level along the $y$-axis for true predicted anomalies. Once again, as $n$ increases, the variance declines and the expected outcome increasingly dominates, causing a drop in true predicted anomalies.

\begin{figure*}[t]
\centering
%\vskip-1.5cm
\includegraphics[width=0.7\hsize]{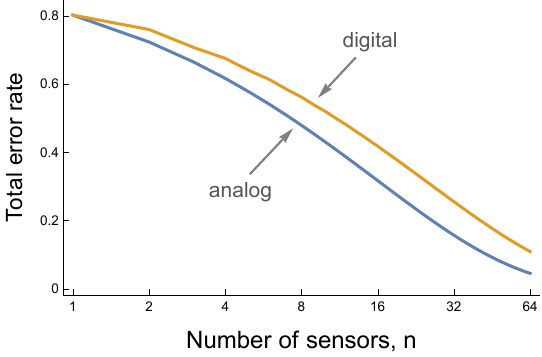}
\vskip0pt
\caption{The cost of digitizing the response of individual sensors. The curves show that an increase in the number of sensors, $n$, reduces the total error rate as the sum of the false negative and false positive rates. In prior figures, the false negative rate is the false predicted anomaly rate, and the false positive rate is one minus the true predicted anomaly rate. The lower blue analog curve corresponds to a circuit that averages the values perceived by the $n$ individual sensors. The upper gold digital curve corresponds to a circuit in which each sensor transforms its input into a $0$ response when the input value is below a sensor-specific threshold and a $1$ response otherwise. For a given number of sensors, $n$, the digital circuit produces more errors because digitization at the individual sensor level loses information.
}
\label{fig:minError}
\end{figure*}

\Figure{minError} shows that analog to digital conversion by sensors decreases the maximum available information. The lower blue curve traces the smaller error rate for a fully analog circuit that averages the actual values coming into the sensors, as in \Fig{oneR}. The upper gold curve shows the rise in the error rate caused by the information lost to digital conversion, as in \Fig{oneR2}.

Digital circuits reduce information but are simpler to construct and often are more robust. Small perturbation will usually not alter the $0/1$ classification by a sensor. By contrast, many sources of noise will cause variability in a measured analog value.

\subsection{Computer network anomaly detection}

In the NSL-KDD dataset of attacks on a central computer, a digital ensemble of sensors performs very well at detecting anomalous computer network characteristics associated with attacks. This dataset is widely used as a benchmark for machine learning studies of anomaly detection. The dataset contains measurements for many features of the computer network \autocite{tavallaee22nsl-kdd}.

A freely available Python notebook calculated how well each of 36 features could independently classify an input as a typical network pattern or an anomalous attack \autocite{dhooge22nsl-kdd-01-eda-oner:}. The analysis used the area under the curve (AUC) to measure the resolving power of a feature, as in \Fig{oneR}. Features with high resolving power included the amount of data sent by the remote computer to the target computer, the amount of data returned to the remote computer, the kind of service request to the target such as email or web page, and the number of recent connections by the same remote computer. 

The AUC values for 22 of 36 features were greater than 0.5, which means those features had some resolving power. The top 15 AUC values ranged from 0.82 to 0.66. If each sensor's response is encoded as $0$ for typical and $1$ for anomalous, then an ensemble digital analysis can be created by summing the values for the 22 resolving features. The ensemble circuit's AUC score is $0.93$, which is good.

F1 provides another measure of classification success, combining how often a positive prediction is correct and how often a positive input is correctly predicted \autocite{powers20evaluation:}. The ensemble circuit's F1 score is $0.9$, which is also good.

Reducing the number of sensors to the top 4 with individual AUC values above 0.75, the ensemble AUC score is $0.92$, and the F1 score is $0.89$. Thus, a small and simple ensemble of digital sensors performs well for this classic benchmark dataset.

\subsection{Extra information in multivariate pattern}

In the ensemble digital model, each sensor passes a digital response. That response can easily be combined with the outputs of other digital sensors to create an overall circuit response. Simple biological circuits may often be built in this way.

\begin{figure*}[t]
\centering
%\vskip-1.5cm
\includegraphics[width=0.75\hsize]{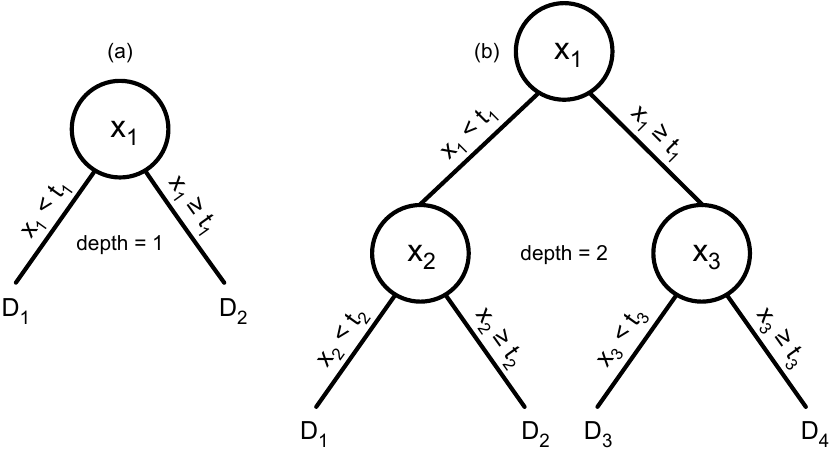}
\vskip0pt
\caption{Decision trees for an anomaly detection classification problem. The challenge is to classify a multivariate data input with values $x_i$ for the $i=1,\dots,N$ data dimensions. (a) A tree of depth 1 that predicts classification based on one feature of a multivariate observation, $x_1$. If $x_1$ is greater than or equal to a threshold, $t_1$, then the tree returns a decision value, $D_2$. Otherwise it returns $D_1$. If there is only a single tree, then the decision value determines the classification. Alternatively, there may be an ensemble of trees, each tree analyzing a different data dimension. In an ensemble, each tree contributes a separate decision value that can be combined with the values from other trees to make an overall classification decision. (b) A tree of depth 2 that uses three different data features.}
\label{fig:simpleTrees}
\end{figure*}

The digital ensemble uses each dimension of the input independently. Each digital sensor takes one input value and responds as a one-step decision tree (\Fig{simpleTrees}a). If the input is greater than some threshold, the decision tree responds one way. Otherwise, it responds the other way.

However, a multivariate pattern rarely occurs as a collection of independent dimensions. Most machine learning methods extract some of the extra multivariate information. The following sections consider two common machine learning circuits that may apply widely in biology.

\section{Boosted decision trees}

\subsection{Deep trees}

A simple extension uses deeper decision trees. In \Fig{simpleTrees}b, the input value for the first feature of the multivariate data is split at some threshold value. If the first feature is greater than its threshold, then a second split is made based on another feature and a different threshold. If the first feature is less than its threshold, then the second split happens based on different criteria.

A deeper tree analyzes multiple inputs, allowing for decisions that include correlations between different feature dimensions of the data. A tree of depth $n$ makes $2^n-1\approx 2^n$ splits on the data. For example, if a system has the capacity to make $2^6=64$ splits, then it can make $2^0=1$ trees of depth 6, or $2^1=2$ trees of depth 5, or $2^2=4$ trees of depth 4, and so on.

Approximately, for $2^n$ splits, the system can make $2^m$ trees of depth $2^{n-m}$. Typically, machine learning applications perform better by using many trees of shallower depth rather than a small number of deep trees. Various methods exist for creating multiple trees and combining them into a single decision ensemble \autocite{zhou12ensemble,hastie17the-elements}.

\subsection{Boosting and biological design}

The most widely successful method creates trees by a boosting process \autocite{schapire13boosting:}. Boosting makes trees sequentially, starting with a single relatively small tree. Then, with an optimized first tree, the algorithm adds a second tree that corrects errors made by the first tree. The process continues adding trees in this way, each tree boosting the success achieved by the prior ensemble.

Boosting seems like a good description of how biological circuits may be designed by natural selection. Initially, a small circuit may provide some information. A second circuit may boost performance, followed by a third, and so on. Sequentially boosted improvement may be the essence of biological design.

\subsection{Typical vs anomalous data as self vs nonself}

\begin{figure*}[t]
\centering
%\vskip-1.5cm
\includegraphics[width=0.98\hsize]{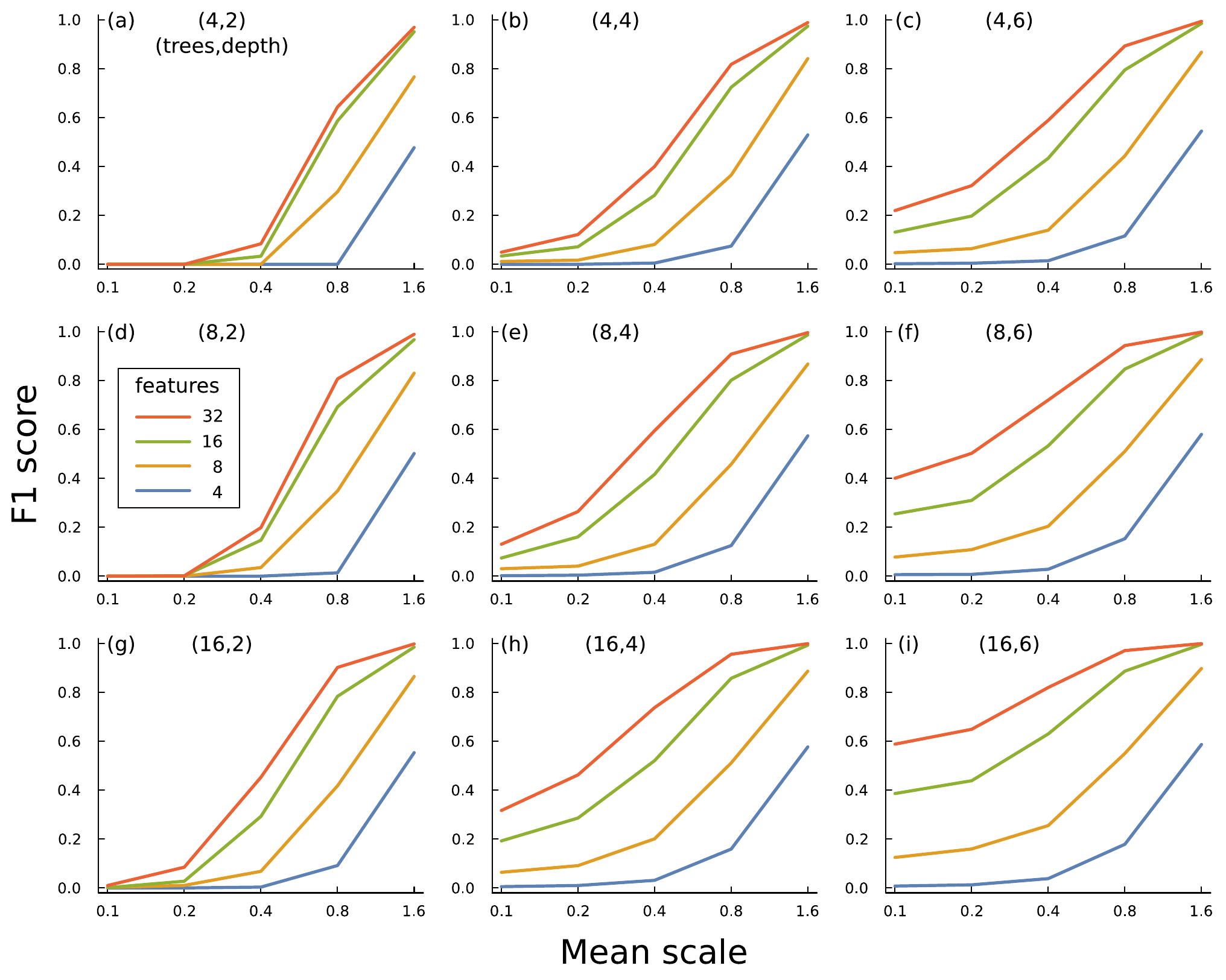}
\vskip0pt
\caption{Performance of boosted decision tree ensembles for classifying typical versus anomalous inputs. \textit{Mean scale} influences the amount of deviation in mean values between typical and anomalous inputs. \textit{F1 score} measures the success of a circuit in classifying typical and anomalous data \autocite{powers20evaluation:}. That score combines how often a prediction of anomaly is correct with how often an anomalous input is correctly identified. \textit{Features} is the number of dimensions in the data. \textit{Trees} is the number of trees in an ensemble circuit. \textit{Depth} is the depth of each tree in a circuit. The text describes the methods and main conclusions for this figure. I generated one dataset with 32 features and used subsets of the feature data for the various plots so that the correlation structure of the data was consistent between the various comparisons. The boosted tree ensemble was calculated by the widely used xgboost algorithm \autocite{chen16xgboost:}. For $T$ trees each of depth $n$, the total number of splits is $\big(2^n-1\big)T$.
}
\label{fig:xgboost}
\end{figure*}

\Figure{xgboost} illustrates some of the tradeoffs in building an ensemble of boosted trees. In this case, I generated an artificial set of data with both typical and anomalous inputs by sampling from multivariate normal distributions. Each input has $f$ feature dimensions. 

For the typical data, each feature dimension has a mean value drawn randomly from a normal distribution with mean zero and standard deviation $\Gs$. I call that standard deviation the \textit{mean scale} because $\Gs$ determines the scale of the fluctuations among the means of the different dimensions.

The variance in each dimension is one, so the $f$-dimensional correlation matrix is also the covariance matrix. I generated that matrix by a random draw from a uniform distribution over all possible correlation matrices \autocite{lewandowski09generating}. Once this distribution is set for typical data, all typical observations come from this single distribution.

For the anomalous data, I used the same process to create a new multivariate normal distribution for each observation. Each anomalous observation is a single random draw from a unique distribution. Thus, classification requires recognizing what a typical observation looks like when compared with a wide variety of anomalous data patterns rather than recognizing specific signatures of danger. This structure captures the essence of self versus nonself discrimination. Here, the typical pattern defines self, and the anomalous observations define nonself, the variety of patterns distinct from self \autocite{aickelin14artificial,dasgupta99artificial,forrest94self-nonself,hofmeyr00architecture,kim01an-evaluation,stibor05a-comparative}.

\subsection{Performance}

\Figure{xgboost}a shows the success of a boosted decision tree circuit. For that panel, the circuit has 4 trees, each of depth 2. As the mean scale increases along the $x$-axis, the circuits do better at detecting anomalies. A greater mean scale implies that, for each feature, the average deviation between the mean values of the typical and anomalous observations rises. Decision trees can easily detect differences in mean values for a feature by splitting at a threshold that likely separates typical and anomalous inputs.

The different curves show the varying numbers of features available in the data. More features tend to increase the largest deviations in mean values between typical and anomalous observations. More features also tend to increase the difference in multivariate correlation structure between typical and anomalous observations because greater dimensionality increases the space of possible correlation patterns.

The other panels show the increase in classification success as the number of trees or the depth of trees increases. Deeper trees are particularly good at identifying differences in multivariate patterns caused by correlations between features. That benefit can be seen by comparing the success of the deeper trees at low values of \textit{mean scale}, for which there is little information available from differences in mean values between typical and anomalous observations.

The structure of this particular problem provides a strong challenge for anomaly detection because no common pattern exists among the anomalous inputs. Additionally, the generating process for the observations creates wide scatter among both typical and anomalous inputs. Nonetheless, the boosted tree ensembles significantly discriminate between typical and anomalous inputs.

\subsection{Boosted trees and biological circuit evolution}

Each node of a tree is simply a binary split based on input. Thus, any biological circuit that expresses the commonly observed Hill response could implement a node of a decision tree \autocite{frank13input-output,zhang13ultrasensitive,martinez-corral24the-hill}. Combining information from multiple trees is also likely to be something that simple biological systems could achieve.

As I mentioned earlier, the sequential process of building boosted trees likely matches the natural tendency for evolutionary processes to create solutions by adding improvements to an initial design. Thus, the simple way in which tree-like decision nodes can be implemented biologically and the sequential process of boosting make boosted trees an excellent model for cellular and neural circuits that solve challenges of classification and decision.

\section{Encoders and internal models}

\subsection{Dimensional reduction}

Encoders reduce dimension by compressing inputs into informative components (for background, see Box 2 of Frank \autocite{frank24circuit}). Dimensional reduction by encoding can be an effective way to identify anomalous environmental conditions. A common autoencoder method first compresses the $f$ features of an input to a representation in a lower $f'$ dimensional space. It then expands that representation back to the original $f$ dimensions, attempting to match closely the original input.

An autoencoder uses patterns in the data \autocite{torabi23practical}. For example, suppose the second feature tends to follow a particular function of the third and fourth features. In that case, the compression method can discard the second feature and recreate that feature during decompression. When a good autoencoder compresses and then decompresses an input, the final decompressed value tends to be close to the original input.

If anomalous inputs lack some of the patterns in typical inputs, an autoencoder built for typical inputs will often distort an anomalous input during the encoding-decoding sequence. The output for an anomalous input will often be farther from the original input than usually happens for typical inputs. Thus, the distance between the input and the output of an autoencoder can be used to classify inputs as typical or anomalous.

Using a sequence of compression steps often creates a more effective encoding. If, for example, the initial data have $2^n$ features, a first compression stage may reduce to $2^{n-1}$ dimensions, followed by compression to $2^{n-2}$ dimensions, and so on. Sequential compression helps to create an internal model of the data \autocite{bengio09learning,bengio13representation,hinton06reducing,ruff21aunifying}. When sequentially compressing images of faces, early steps may focus on facets such as eyes, ears, nose, and mouth. Later steps may consider relations between those parts, providing an internal model of how a typical face tends to look \autocite{le12building,lee09convolutional,masci11stacked}.

\subsection{Anomaly detection}

For this article, we are particularly interested in the simplest effective circuits. A full autoencoder requires both encoding compression and decoding decompression. A simpler approach uses only the encoding step. We convert the $f$ features in the input to $f'$ compressed dimensions. If we design an encoder that tends to make a large distance between typical and anomalous inputs in the $f'$ dimensional representation, then we can use that distance to detect anomalies.

\begin{figure*}[t]
\centering
%\vskip-1.5cm
\includegraphics[width=0.8\hsize]{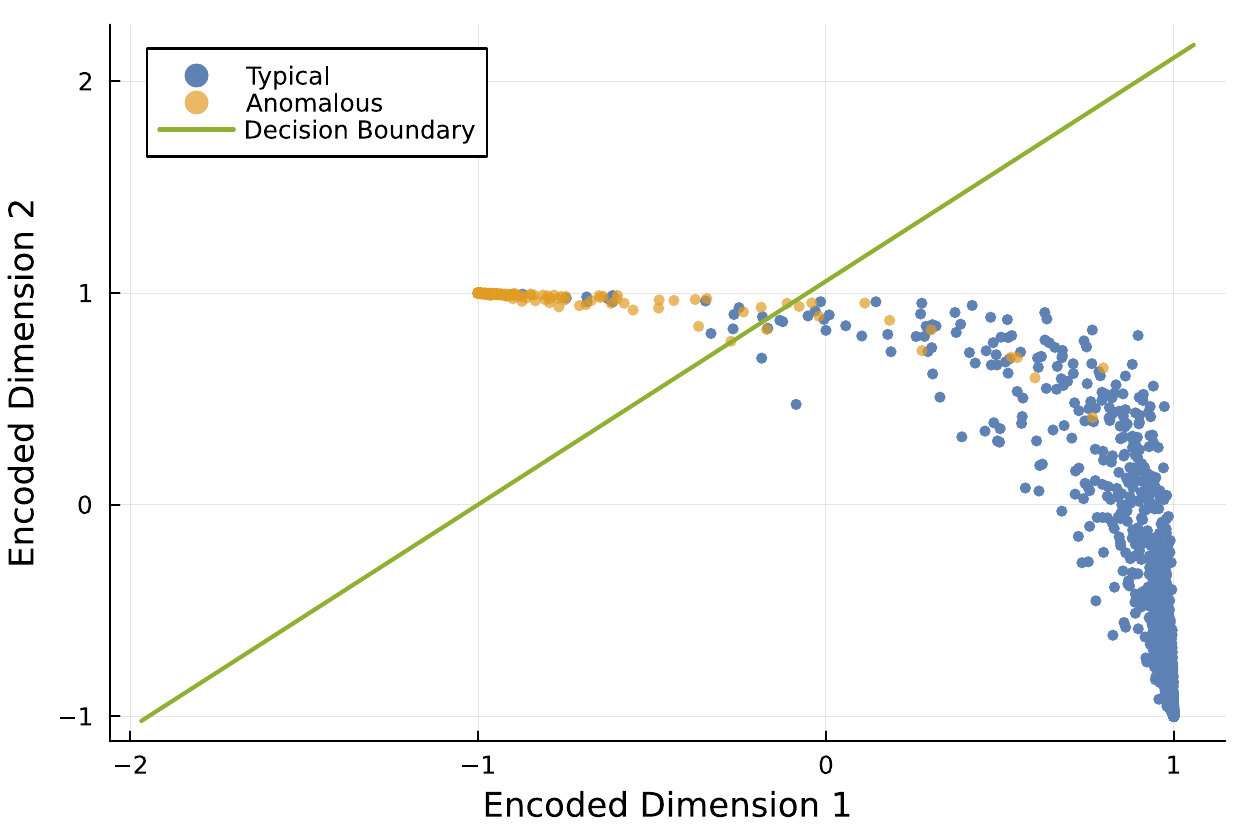}
\vskip0pt
\caption{Encoder model that reduces 4-dimensional inputs to 2 dimensions, separating typical and anomalous observations. I used the same methods to generate the data as for boosted trees, described previously. Of the initial $100,000$ observations, $10\%$ are anomalies, and the rest are typical. I randomly split the data into a training set comprised of $70\%$ of the observations and the remainder in the test set to evaluate the fitted model. This plot shows a random subset of the test data with approximately $2700$ typical observations and $300$ anomalous observations. Compared with \Fig{encoder1}, the mean scale value here is 1.6, and the number of features is 4. I used a distinct dataset for this figure to provide a visualization that shows the separation between typical and anomalous points more clearly. In this case, the F1 score is $0.96$, corresponding to relatively few misclassified points. The model encoded the 4 input dimensions to 2 output dimensions with a single layer of a neural network, using 10 parameters.
}
\label{fig:encoderScatter}
\end{figure*}

\Figure{encoderScatter} illustrates how an encoder separates typical and anomalous inputs. In this example, the 4 input dimensions were reduced to 2 output dimensions using a single layer neural network. That small network separated typical and anomalous observations with high accuracy.

\subsection{Factors influencing circuit accuracy}

\begin{figure*}[t]
\centering
%\vskip-1.5cm
\includegraphics[width=0.49\hsize]{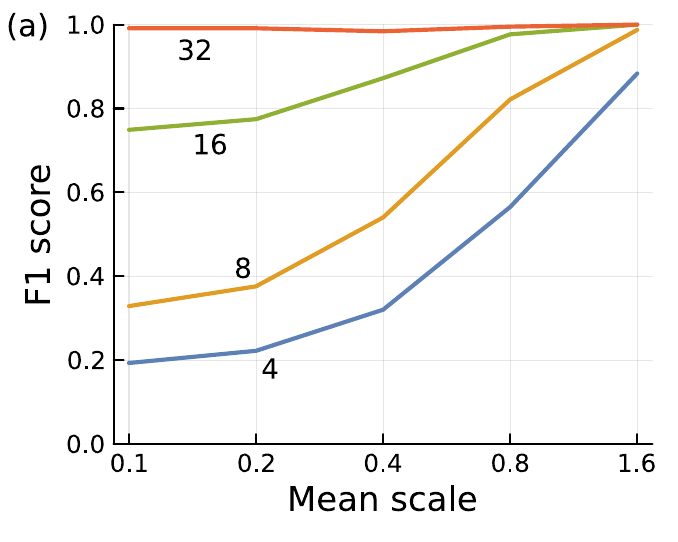}
\includegraphics[width=0.49\hsize]{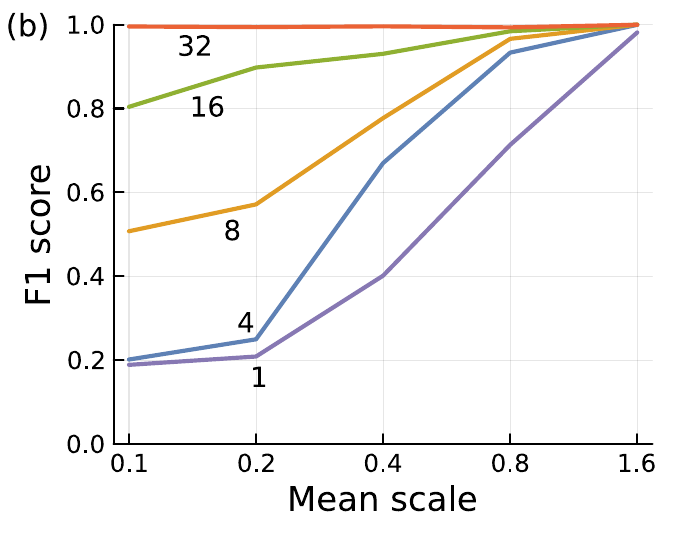}
\vskip0pt
\caption{Encoder model to separate typical from anomalous inputs. The labels on each curve describe the number of features, $f$, in the data. I used the same methods to generate the data as for boosted trees, described previously. (a) I generated three separate input datasets and calculated F1 scores for each to compensate for peculiarities of any particular dataset. I then averaged the three values for each mean scale by feature combination. The overall pattern and magnitudes for each separate dataset were similar. For each dataset, I generated 32 features. I then used the first $f$ features in the set for each curve. If we compress from inputs with $f=2^n$ feature dimensions to a single output dimension, then a full encoder model has $(2f+5)(f-1)/3$ parameters when $n$ is an integer. (b) I began calculation for each point using all 32 features. I then iteratively deleted the single feature that provided the least information, measured by the smallest reduction in F1 when deleting that feature. I continued until the specified number of features for a particular curve remained. This iterative deletion method provided a better set of features than simply picking the first $f$ features as in (a).
}
\label{fig:encoder1}
\end{figure*}

\Figure{encoder1}a compares an encoder's classification efficacy under different conditions. The F1 score measures classification efficacy, combining how often a positive prediction is correct and how often a positive input is correctly predicted \autocite{powers20evaluation:}. The mean scale influences the amount of variation between typical and anomalous mean values in each dimension of the data.

The full data initially contained $f=32$ features. I then calculated F1 scores by using only the first $f=4,8,16$ feature dimensions. Each line in the figure is labeled with the number of features used, $f=2^n$. For this figure, an encoder reduces dimensionality from the $f=2^n$ inputs to $2^0=1$ output, using $n$ layers in the neural network encoder.

Three conclusions follow from this figure. First, between typical and anomalous inputs, bigger differences in mean values for each dimension make it easier to detect anomalies, shown in the figure as the mean scale increases along the $x$-axis.

Second, sampling more features of the data improves classification. The improvement occurs primarily for small values of the mean scale, in which mean differences provide limited information. In those situations, a classifier can succeed when it is able to infer distinctions between typical and anomalous inputs in the correlation pattern among the dimensions. In this example, raising the number of features enhances the information about correlational pattern, providing increasingly accurate classification.

The third conclusion is that, given a sufficient number of features, an encoder can achieve nearly perfect classification for these input data. In this case, an encoder using all 32 features of the data made very few classification errors.

The encoder for $f=32$ features achieved high success by optimizing the 713 parameters in its encoding network. For $f=4,8,16$, the circuits required 13, 49, and 185 parameters.

\subsection{Simplifying circuits}

\Figure{encoder1}b shows that, for a given performance level measured by F1, we can find simpler circuits with the same performance. In that plot, the calculation of each point began with all 32 features. I then iteratively removed one feature at a time, dropping the feature that provided the least amount of information, measured by the smallest decline in F1. I continued dropping features in this way, providing an F1 measure for $f=1,2,\dots,32$ for each mean scale level. The plot shows curves for particular $f$ values.

Choosing the best $f$ features of the full 32 in the data provides a better F1 score than taking the first $f$ features in the data, as expected. Put another way, for a given F1 level of performance, we can use a smaller circuit if we select the best features rather than using a fixed feature set. The amount by which a circuit can be reduced for a given F1 score depends on the particular data structure, as shown by the plots.

We could further reduce the number of parameters in a circuit by imposing a cost on each parameter. A cost favors a parameter to decline close to zero when it adds relatively little improvement in performance. We then obtain a simplified circuit by pruning all parameters near zero.

Overall, relatively small encoder circuits can achieve good classification for some types of data.

\section{Discussion}

I have focused on anomalies as unusual observations, anything that differs from what is typical. Detection does not depend on specific anomalous patterns or danger signals. Instead, a system creates a model of a typical input and infers when an input differs from that internal model. Something like ``That's an unusual smell'' or ``I've never seen that before.''

Sensory or neural adaptation provides a simple example. Many biological circuits adjust their baseline by averaging over recent inputs. That baseline allows the circuit to measure deviations from what has recently been typical \autocite{carandini12normalization,fairhall01efficiency,kohn07visual,wark07sensory,whitmire16adaptation}. I presented simple circuits in \Eqq{anomtrend} and \ref{eq:foldchange} that adapt to recent trends. An ensemble of such circuits could classify multivariate inputs.

Self versus nonself recognition occurs widely throughout biology \autocite{bedinger17recognition,boehm06quality,boehm21self/nonself,cooper10self/non-self,de-tomaso18the-evolution,mojica16the-discovery,pradeu13immunity,ruhe13bacterial,tock05the-biology,witzany16self-nonself}. In some cases, nonself is recognized by direct pattern recognition, which does not require the more challenging kinds of circuits discussed in this article. In other cases, self recognition is more complex and not fully understood \autocite{koncz24a-journey}. It seems that systems sometimes recognize what is self and classify as anomalous those observations that do not fit the self pattern, potentially sharing properties with the machine-learning circuits discussed in this article.

The human hippocampus appears to recognize novelty in certain contexts \autocite{kumaran06anunexpected,kumaran07match--mismatch,bhasin22dynamic}. Further studies suggest that memory creates a model of what is common. The system classifies inputs as novel or unusual when they deviate significantly from expectation \autocite{frank21expectation-driven,shing23prediction}. With regard to the analyses in this article, some sort of dimensional reduction likely encodes the internal model.

Cellular and physiological systems would likely gain from anomaly detection. The models in this article suggest the kinds of small circuits that could work within these constrained biological systems.

\section*{Acknowledgments}

\noindent The Donald Bren Foundation, US Department of Defense grant W911NF2010227, and US National Science Foundation grant DEB-2325755 support my research.

\section*{Data availability statement}

Software to produce the figures is available on GitHub \autocite{frank24circuit-codeb}.

%\vfill\eject

\mybiblio	% uses main.bib by default, add other bibs as needed

% used cuted package strip env to force balancing of columns
%\ifmulticol\begin{strip}\hbox{\null}\end{strip}\hbox{\null}\fi

\end{document}

%% file: ml-def.tex
\newcommand*{\Gg}{\gamma}

\newcommand*{\Gl}{\lambda}

\newcommand*{\Gs}{\sigma}
\newcommand*{\Gt}{\tau}

\newcommand*{\Gf}{\phi}

\usepackage{bm}

\DeclarePairedDelimiter\abs{\lvert}{\rvert}
\DeclarePairedDelimiter\norm{\lVert}{\rVert}
\DeclarePairedDelimiter\angb{\langle}{\rangle}
\DeclarePairedDelimiter\lrb{\lbrack}{\rbrack}
\DeclarePairedDelimiter\lr{\lparen}{\rparen}
\DeclarePairedDelimiter\lrbr{\lbrace}{\rbrace}

\makeatletter
\let\oldabs\abs \def\abs{\@ifstar{\oldabs}{\oldabs*}}
\let\oldnorm\norm \def\norm{\@ifstar{\oldnorm}{\oldnorm*}}
\let\oldangb\angb \def\angb{\@ifstar{\oldangb}{\oldangb*}}
\let\oldlrb\lrb \def\lrb{\@ifstar{\oldlrb}{\oldlrb*}}
\let\oldlr\lr \def\lr{\@ifstar{\oldlr}{\oldlr*}}
\let\oldlrbr\lrbr \def\lrbr{\@ifstar{\oldlrbr}{\oldlrbr*}}
\makeatother

\newcommand*{\dd}{\textrm{d}}
\newcommand*{\Eq}[1]{eqn~\ref{eq:#1}}
\newcommand*{\Eqq}[1]{eqns~\ref{eq:#1}}

\newcommand*{\Figure}[1]{Figure~\ref{fig:#1}}
\newcommand*{\Fig}[1]{Fig.~\ref{fig:#1}}

\newcommand*{\boldrule}{\hrule height 1.2pt}
\newcommand*{\noterule}{\medskip\boldrule\medskip}	% for notes